# Transport in an inhomogeneous interacting one–dimensional system

I. Safi and H.J. Schulz

*Laboratoire de Physique des Solides, Université Paris–Sud, 91405 Orsay, France*

## Abstract

Transport through a one–dimensional wire of interacting electrons connected to semi–infinite leads is investigated using a bosonization approach. An incident electron is transmitted as a sequence of partial charges. The dc conductance is found to be entirely determined by the properties of the leads. The dynamic nonlocal conductivity is rigorously expressed in terms of the transmission. For abrupt variations of the interaction parameters at the junctions the central wire acts as a Fabry–Perot resonator. When one of the connected wires has a tendency towards superconducting order, partial Andreev reflection of an incident electron occurs.

72.10.–d, 73.40.Jn, 74.80.Fp

Typeset using REVTEX

Due to advances in semiconductor microtechnology it is now possible to fabricate high mobility quantum wires which are very close to being ideally one–dimensional.[1,2] In one–dimensional electron systems interactions play a crucial role, giving rise to the so–called Luttinger liquid behavior.[3,4] Transport measurements are a possible method to test theoretical predictions. For example, applying an electric field over a finite segment of a homogeneous infinitely long wire, Kane and Fisher[5] find (for spinless electrons) a conductance $g = g_0 K$, where $g_0 = e^2/h$ is the conductance quantum and $K$ is a nonuniversal number depending on the interactions in the wire. However, it seems more reasonable to describe a two–probe measurement by connecting the interacting wire to two long leads intended to mimic the role of reservoirs where the chemical potential is fixed. This type of idealization was often used in theoretical work on transport in mesoscopic devices.[6–8] This work continues on the path paved by Landauer[9] who relates the conductance of a coherent device to the transmission $T$ of an incident electron on the device viewed as a scattering entity. Depending on assumptions about the measuring procedure[7] one finds either[6]

$$g = g_0 T \qquad (1)$$

or Landauer's original result,[9,10] $g = g_0 T/(1-T)$ ($T$ is the transmission coefficient). Eq.(1) agrees with the conductance of a one–channel ballistic constriction for which $T = 1$.[11] In Landauer–type theories, interactions are mostly ignored,[12] or at least are accounted for by a phenomenological time larger than the characteristic time to cross the sample. It's our purpose to include interactions explicitly and to investigate the transmission processes through a finite wire as well as the effect of the contacts on its transport properties. Confined to one–dimensional sample and leads, our electrons are interacting everywhere, but in general with different strength in the leads and in the sample. We will not consider disorder, and deal with the non–local time or frequency dependent conductivity, containing much more information than the bulk conductance. Also, due to the space–dependent interactions phenomena familiar from metal–superconductor contacts like Andreev reflection[13] and the proximity effect will be seen.

Momentum–conserving interactions between spinless electrons can be described by the bosonized Hamiltonian[3,4]

$$H = \int_{-L}^{L} \frac{dx}{2\pi} \left[ uK(\partial_x \Theta)^2 + \frac{u}{K}(\partial_x \Phi)^2 \right] \quad (2)$$

where the boson field $\Phi$ is related to the particle density by $\rho - \rho_0 = -\partial_x \Phi/\pi$, and $\partial_x \Theta/\pi$ is the momentum conjugate field to $\Phi$. The interaction–dependent parameter $u$ determines the velocity of the elementary excitations, and $K$ (also interaction–dependent) determines the algebraic decay of correlation functions, indicating a tendency of the Luttinger liquid towards either superconducting or CDW order, depending on whether $K > 1$ or $K < 1$. A "Fermi liquid" is found for the noninteracting case, $K = 1$, $u = v_F$. We now consider a finite interacting wire perfectly connected to two identical leads at its end points $\pm a$. We shall label the quantities pertaining to the leads (central wire) by the subscript 1 (2). In $H$ the parameters $u, K$ then vary from $u_2, K_2$ in the wire to $u_1, K_1$ outside. We use periodic boundary conditions, i.e. we join the exterior wires to form a ring of length $2L$. This simulates two semi–infinite perfect leads if we put $K_1 = 1$ and $L \gg a$. Thus the electrons coming out the central wire can not come back to it, the interactions being absent and the time to go around the ring too long. We can therefore define properly the reflection and transmission coefficients of an electron incident on the central wire. Creating an electron amounts to introducing a kink of height $\pi$ in $\Phi$, and the problem of reflection and transmission then reduces to a solution of the equations of motion for $\Phi$. This leads us naturally to diagonalize the Hamiltonian (2) with space–dependent $u$ and $K$: we expand the field $\Phi$ in terms of a discrete set of boson creation and annihilation operators and eigenfunctions.

We start with the case where $u$ and $K$ jump from $u_1, K_1 = 1$ to $u_2, K_2$ at $\pm a$. We call $t_y$ the time it takes for an electron to go from $y$ on the lead to the closest contact, i.e, $u_1 t_y = |y| - a$, and $t_2 = 2a/u_2$ is the traversal time of the central wire. $\Phi$ obeys simply a wave equation with velocity $u_1$ outside and $u_2$ inside. The propagating solutions have to be joined at $\pm a$. The system of coupled equations of motion for $\Phi, \Theta$ require their continuity, and so is $\frac{u}{K}\partial_x \Phi = \partial_t \Theta$. Indeed, a discontinuous $\Phi$ would lead to an unphysical singularity:

a charge accumulation, thus a nonconservation of the current $j = \partial_t \Phi/\pi$, as one can derive from the continuity equation. Note that the continuity of $\Phi$ and $\Theta$ guarantees that of the fermion field. In the leads, the propagating solutions can be simply related to the original right and left going electrons density $\rho_\pm$. In the interacting wire we instead define their counterpart by

$$\tilde{\rho}_\pm = \frac{1}{2}(\rho \pm j/u) \qquad (3)$$

For the sequel, it's useful to write the time evolution of their corresponding current $j_\pm(x) = \pm u \tilde{\rho}_\pm(x)$ (defined for any $x$) as:

$$J(x,t) = \int_{-L}^{L} \frac{dy}{u} M(x,y,t) J(y,0) \qquad (4)$$

where $J = (j_+, j_-)$ and $M$ is a $2 \times 2$ integral kernel matrix which can be simply interpreted for $a \leq |x|, |y| < L$: $M_{rr'}(x,y,t)$ $(r,r' = \pm 1)$ is the electronic charge with velocity $r'u_1$ reaching $x$ at time $t$ if an electron initially emanates at $y$ with velocity $ru_1$ (i.e. if $\langle \rho(x) \rangle = r \langle j(x) \rangle / u_1 = \delta(x-y)$ at $t = 0$). $M$ verifies time reversal symmetry: $M_{rr'}(t) = M^*_{-r-r'}(-t) = M_{-r-r'}(-t)$; we'll henceforth restrict $t$ to $[0, t_L]$. For points on opposite leads verifying $a \leq -y, x < (L-a)/2$, all the entries of $M$ but $M_{++}$ are zero: this expresses the fact that only the direct path allows a particle to travel from $y$ to $x$, the other trips would take more than $t_L$. Therefore, there is no ambiguity in identifying $M_{++}(x,y,t)$ with the charge transmitted from $y$ to $x$ at time $t$. For points on the same lead, e.g. the left one, $M_{+-}(-L < y', y \leq -a, t) = 0$: a left–going electron originating at $y$ will travel counterclockwise and takes at least $4t_L - t_{y'} - t_y$ to generate a right–going charge at $y'$ after its reflection on $a$. Besides, $M_{rr}(y', y, t) = \delta[(y'-y)/u_1 - rt]$ (direct propagation) while $M_{-+}(y', y, t)$ can be identified with the reflected charge appearing at $(y', t)$ for an initial right going electron at $y$.

Our simple model allows us to express all the $M_{rr'}$ (for any $x, y$ and for times less than $t_L$) solely in terms of the "descending Dirac comb" $\Delta(t) = \sum_{0 \leq p < \frac{t_L}{t_2}} \gamma^{2p} \delta(t - 2pt_2)$, where $\gamma = (1 - K_2)/(1 + K_2)$ is the reflection coefficient of an electron incident on the contact between wires 1 and 2. $\Delta(t)$ is nothing but $M_{++}(x,x,t)$ for $|x| \leq a$: it reveals the cyclic motion inside the finite wire. A peak in $\langle \tilde{\rho}_+ \rangle$ initially localized at $x \in [-a,a]$ reappears at $x$ after each time elapse $2t_2$, with its charge reduced by $\gamma^2$ due to two successive reflections at the end points.

If an electron emanates initially at $y \leq -a$, the charge transmitted to $x \geq a$ at time $t < t_L$ is

$$M_{++}(x,y,t) = (1-\gamma^2)\Delta\left[t - (t_x + t_y + t_2)\right] \tag{5}$$

The first Dirac peak, occurring at $t_0 = t_x + t_y + t_2$ accounts for the first transmission to $x$ of the partial charge $1 - \gamma^2$, after two reflections at $-a, a$ on the way. Each subsequent transmission at $t = t_0 + 2pt_2$ is reduced by $\gamma^{2p}$, i.e. *the incident electron is into a series of non–integer charged maxima in the charge density.* The series sums up to unity in the limit $t_L \gg t_2$: *the transmission is perfect.* It would be cumbersome to write out $M$ for other locations. Let's just add that the expression of $M_{+-}(y' \leq -a, y \leq -a, t)$ reveals a first reflection with coefficient $\gamma$, while the subsequent ones are of opposite sign and sum up to $-\gamma$, leading to a vanishing total reflection.

The nonlocal dynamic conductivity $\sigma(x,y,t)$ is given by

$$\sigma(x,y,t) = g_0 K(y)\theta(t) \sum_{r,r'} M_{rr'}(x,y,t) \tag{6}$$

To obtain this result one has just to realize that $\sigma(x,y,t)$ yields exactly $\langle j(x,t) \rangle$ in response to a electric field pulse $\delta(t)E$ applied at $y$. This field generates in turn an initial current peak at $y$ so that $\langle j(x,t) \rangle$ develops in accordance with the right hand side of eq.(6) (see eqs.(3) and (4)). For points on opposite leads ($x \geq a$, $y \leq -a$) and $t < t_L$, eq.(6) reduces to

$$\sigma(x,y,t) = g_0 M_{++}(x,y,t) \tag{7}$$

This is a generalization of the Landauer formula, eq.(1), to a dynamic situation in the sense that it relates transport properties to transmission. To infer the frequency dependence of the conductivity, caution is needed in the order of limits: $t_L$ is taken to infinity before $\delta^{-1}$, the adiabatic turn–on time of the dc electric field: we neglect $e^{-\delta t_L} \ll 1$, thus preventing the electrons and holes generated by the electric field to go all around. With $\overline{\omega} = \omega + i\delta$, from eq.(7) we find $\sigma$ for points on opposite sides as

$$\sigma(x, y, \overline{\omega}) = g_0 \exp i\overline{\omega}(t_x + t_y) \frac{1 - \gamma^2}{\exp(-i\overline{\omega}t_2) - \gamma^2 \exp(i\overline{\omega}t_2)}. \tag{8}$$

For points inside $[-a, a]$, all the entries of $M$ contribute and we obtain

$$\sigma(x, y, \overline{\omega}) = g_0 K_2 \left\{ \exp \frac{i\overline{\omega}}{u_2} |x - y| \right.$$
$$\left. + \frac{\gamma}{\exp(-2i\overline{\omega}t_2) - \gamma^2} \sum_{r=\pm 1} \left[ \gamma \exp \left[ \frac{i\overline{\omega}r}{u_2}(x - y) \right] + \exp \left[ \frac{i\overline{\omega}}{u_2}(r(x + y) - 2a) \right] \right] \right\}. \tag{9}$$

We can deduce $\sigma(-a, a, \overline{\omega})$ from both expressions (the current being continuous at the junctions), and it's easier to see from eq.(8) that it reaches its maximum $g_0$ at each eigenfrequency of the central wire $\omega_n = 2n\pi/t_2$ (this is so if we keep $t_2$ finite, while $\delta \to 0$). This is the resonant absorption of an electric field by a finite system. The usual Dirac peaks are broadened by the connection to infinite leads as shown in the figure.

In summary, the finite wire behaves as a Fabry–Perot resonator with the junctions playing the role of symmetric mirrors. One can ask which results persist if we consider the more general situation where $u$ and $K$ vary inside the wire and can be asymmetric. We also allow for interactions in the leads with constant $u_1$ and $K_1$. Following the previous steps, we can still show that for points on opposite leads, only the matrix element $M_{rr}$ is nonzero, $r$ being the sign of the direct way to go from one point to the other. Even if the leads are interacting, we can continue to view $M_{++}(x, y, t)$ as the transmitted charge at $(x > a, t)$ when a unit flux was incident at $(y \leq -a, 0)$. This is because wherever $\langle \widetilde{\rho}_- \rangle$ vanishes, $\langle j \rangle = u_1 \langle \rho \rangle = u_1 \langle \widetilde{\rho}_+ \rangle$ so that a peak in the total density propagates at velocity $u_1$. Similarly, the reflection is expressed through $M_{-+}$. Denoting $\lim_{\omega \to 0} F(x, y, \omega + i\delta) \equiv F(x, y)$ where $F$ is any function, we get $M_{++}(x \geq a, y \leq -a) = 1$ and $M_{-+}(y', y) = 0$. As before, the total transmission through the central wire is perfect, even if $u, K$ are not symmetric on $[-a, a]$. This fact is not surprising: both $j$ and $u/K\rho$ are uniform in the steady state. But $u/K$ is the same on the opposite leads, and so is $\langle \tilde{\rho}_+(x, \omega = 0) \rangle$. A partial transmission can occur only if the leads are asymmetric.

We can also generalize the decomposition (6) which relies merely on the definition of $\sigma$ and $M$ and may also be checked from their eigenfunction expansion. If $y < -a$, one has just to replace $K(y)$ by $K_1$ instead of 1, so that the identity (6) yields, for points on the leads: $\sigma(x \geq a, y \leq -a, t) = g_0 K_1 M_{++}(x, y, t)$ and $\sigma(y' \leq -a, y \leq -a, t) = g_0 K_1 [\delta(t - |y' - y|/u_1) + M_{-+}(y', y, t)]$. In the zero–frequency limit the transmission is perfect: $\sigma(x, y) = \sigma(y', y) \equiv g_0 K_1$. This is also true if one or both of $x$ and $y$ are in the central wire because $\sigma(x, y)$ is independent on its arguments as we can check from its mode expansion. Actually, this is a constraint to be obeyed by any sensible transport theory for any one–dimensional system exhibiting time reversal symmetry.[14] We can verify explicitly this constraint in the special model we solved ($K = K_2$ on the central wire) by taking the limit $\omega \to 0$ leaving $t_2$ finite. Our results (eqs.(5), (8)) concern $K_1 = 1$, but we can rewrite them for $K_1 \neq 1$ provided the reflection coefficient $\gamma$ is replaced by $\gamma' = (K_1 - K_2)/(K_1 + K_2)$. We thus recover $\sigma(x, y) = g_0 K_1$ for any $x, y$, even if these are on different wires.

Let's discuss the conductance one can measure. If we impose a current through a non–dissipative system, no voltage drop is measured and the conductance would be infinite. Conversely, we can connect it to perfect leads intended to simulate reservoirs where dissipation takes place and where the chemical potential is not affected by the current through the sample.[7] We define the conductance $g$ of the central wire as the ratio of the current to the potential drop between the junctions imposed by the leads. The general relation of the current to the electric field then shows that $g$ is given by the uniform value of $\sigma(x, y) = g$, where $x, y$ can be inside or outside $[-a, a]$. Therefore, $g = g_o K_1$ is the conductance of the central wire, irrespective of the form of its own parameters.

Returning to the initial model with abrupt variations of $u$ and $K$, it's quite instructive to take a different limit where $t_2$ is of the same order as $t_L$. In particular, $t_2 \gg \delta^{-1}$ so that we can neglect $e^{-\delta t_2}$ in the expressions of $\sigma$. For instance, if $x, y$ are in $[-a, a]$, eq.(9) becomes (after $\gamma \to \gamma'$)

$$\sigma(x,y,\overline{\omega}) = g_o K_2 \left\{ \exp[i\overline{\omega}|x-y|/u_2] + \sum_{r=\pm} \gamma' \exp[i\overline{\omega}(t_2 + r(x+y)/u_2)] \right\} . \qquad (10)$$

We focus on three regions, corresponding to the neighborhood of the origin (the "bulk"), denoted by $N_b$ ($|x| \ll a$), and to that of each contact, denoted by $N_\pm$ ($|x \pm a| \ll a$). The second exponentials are of order $e^{-\delta t_2}$ in $N_b$, thus $\sigma$ regains its bulk form in an homogeneous infinite wire. In $N_\pm$, $\sigma$ depends on both $x$ and $y$, but the zero frequency limit is uniform on $N_\pm$: $\sigma(x,y) \equiv 2g_o/(K_1^{-1} + K_2^{-1})$.[15]

If we can do a measurement which doesn't introduce any additional scattering mechanism and restrict an electric field to a segment in one of those neighborhoods, their respective conductances are $g_b = g_0 K_2$ and $g_\pm = 2g_o/(K_1^{-1} + K_2^{-1})$. $g_b$ is the value of the bulk conductance as derived in ref. 5. $g_\pm$ can be thought as a conductance of one junction: it's worth noting that $g_\pm$ coincides with the transmission through $\pm a$, multiplied by the ratio $u_1/u_2$.

In the same limit of large $t_2$, the local correlation functions for superconducting pairing decay asymptotically in time as $\tau^{-2\kappa}$ with $\kappa_b = 1/K_2$ near the origin (as is known for an homogeneous wire), while $\kappa_\pm = 2/(K_1 + K_2)$ near the junction points. If $K_1 = 1$ and $K_2 > 1$, $\kappa_b < \kappa_\pm < 1$: the pairing fluctuations in the central wire extend to the external leads. This is the analogue of a proximity effect for the situation where there are only power law pairing correlations but no long–range order. On the other hand, an electron injected in one lead and incident on the junction is reflected with coefficient $\gamma = (1 - K_2)/(1 + K_2) < 0$, i.e. *a partial hole is reflected back*. We recognize the phenomenon of Andreev reflection:[13] when an electron is incident on a normal metal–superconductor interface, it needs to make a pair with an electron from the normal metal to enter the superconductor. According to whether its energy is lower or higher than the superconducting gap, an entire or partial hole is reflected back. In the present case there is no gap, so we only get a partial hole reflected. However, in the limit $K_2 \to \infty$, we get exactly one hole reflected.

To summarize, we have investigated transmission and transport through a finite clean interacting wire perfectly connected to semi–infinite one–dimensional leads, all being treated on the same footing. The total transmission of an incident flux on the finite wire is perfect: this is related to the momentum conserving interactions and the symmetry of the external leads. Accordingly, the conductance one measures is just $g_o K_1$, regardless of the internal $u, K$. If $u, K$ are also constant on the central wire, the ac conductivity linking the end points shows a resonance at the eigenfrequencies of the finite wire. The multiple reflections processes at the contacts, as in a Fabry–Perot resonator, shed light on the perfect total transmission and the departure of the conductance from its bulk value $g_0 K_2$ to $g_0 K_1$. This is another manifestation of the fact that the measured conductance is sensitive to the geometry and may be governed by the scattering at the contact probes. In real experiments, inhomogeneities at the contacts are unavoidable and the device opens into wide two dimensional electrodes. The effects of this deserve further study. Nevertheless, the experimental results[2] obtained in quantum wires tend to confirm the relevance of the $g = g_o$ result.

## FIGURES

FIG. 1. The real part of the the nonlocal conductivity through the contacts versus the ratio of the external frequency to the proper frequency of the wire $2\pi/t_2$ for $|\gamma| = 0.1$ (solid line), $|\gamma| = 0.5$ (dotted line) and $|\gamma| = 0.9$ (dash–dotted line). If $K_1 \neq 1$, $\sigma$ is multiplied by $K_1$ and $\gamma$ is replaced by $\gamma'$. The resonance (respectively antiresonance) at integer (respectively half–integer) values correspond to symmetric (respectively antisymmetric) modes of the central wire.

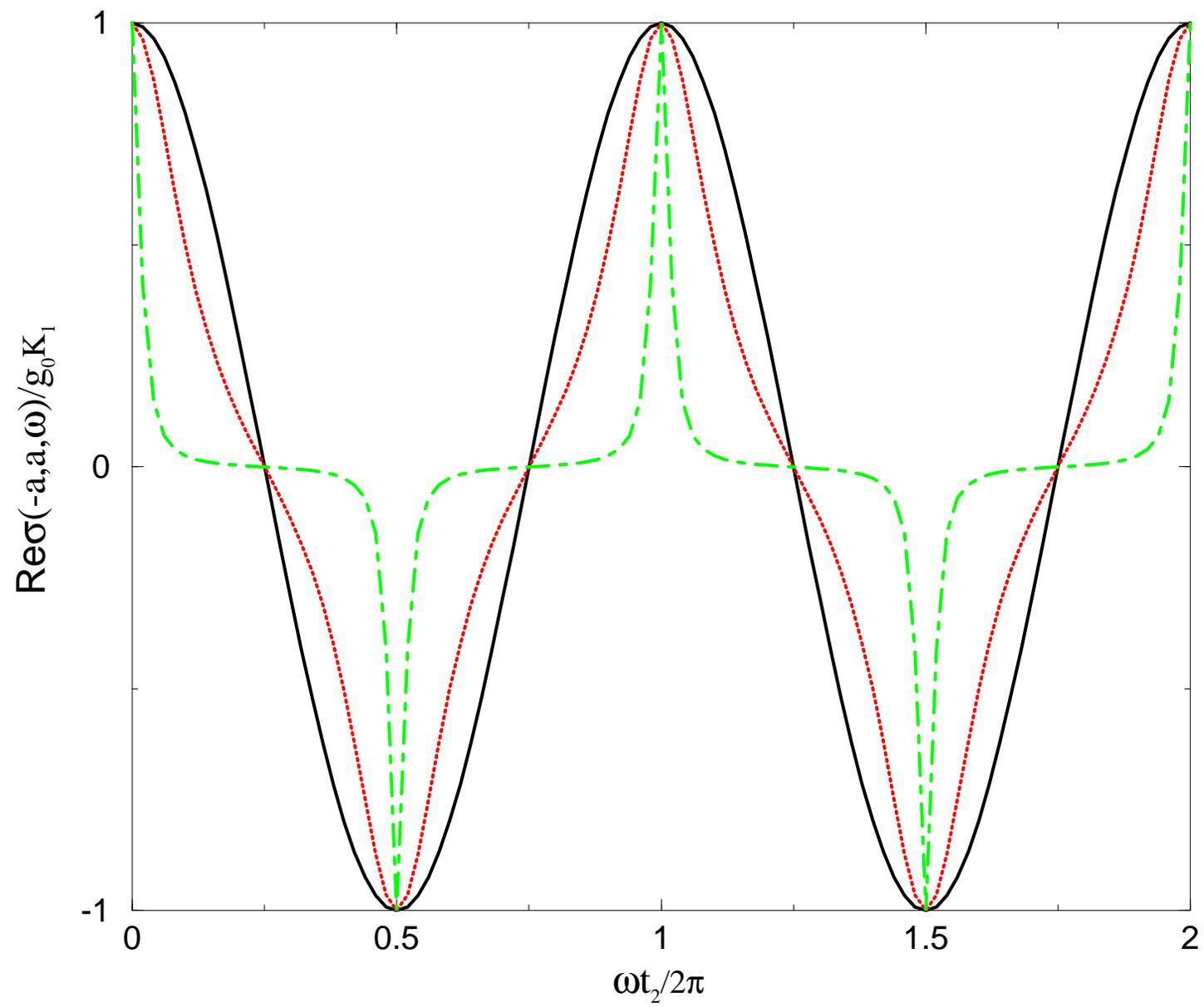